\documentclass[%
    reprint,
    aps,
    physrev,
]{revtex4-2}
\usepackage{graphicx}
\usepackage{microtype}              
\usepackage[utf8]{inputenc}         
\usepackage[T1]{fontenc}            
\usepackage{newtxtext}              
\usepackage[smallerops]{newtxmath}  
\usepackage{nicefrac}
\usepackage[dvipsnames]{xcolor}
\usepackage{hyperref}
\hypersetup{
    colorlinks=true,
    linkcolor=NavyBlue,
    citecolor=NavyBlue,
    urlcolor=NavyBlue
}

\DeclareMathOperator{\D}{D}
\DeclareMathOperator{\KL}{KL}
\DeclareMathOperator{\tr}{tr}
\DeclareMathOperator{\diag}{diag}
\newcommand*{\tran}{^{\mkern-1.5mu\mathsf{T}}}
\newcommand{\renyi}{R{\'e}nyi}
\begin{document}\frenchspacing
\title{Generalized Information Bottleneck for Gaussian Variables}
\author{Vudtiwat Ngampruetikorn}
\author{David J.~Schwab}
\affiliation{%
    Initiative for the Theoretical Sciences,
    The Graduate Center, CUNY, 
    New York, NY 10016, USA%
}
\begin{abstract}
The information bottleneck (IB) method offers an attractive framework for understanding representation learning, however its applications are often limited by its computational intractability. 
Analytical characterization of the IB method is not only of practical interest, but it can also lead to new insights into learning phenomena. 
Here we consider a generalized IB problem, in which the mutual information in the original IB method is replaced by correlation measures based on \renyi\ and Jeffreys divergences. 
We derive an exact analytical IB solution for the case of Gaussian correlated variables. 
Our analysis reveals a series of structural transitions, similar to those previously observed in the original IB case. 
We find further that although solving the original, \renyi\ and Jeffreys IB problems yields different representations in general, the structural transitions occur at the same critical tradeoff parameters, and the \renyi\ and Jeffreys IB solutions perform well under the original IB objective.
Our results suggest that formulating the IB method with alternative correlation measures could offer a strategy for obtaining an approximate solution to the original IB problem. 
\end{abstract}
\maketitle
\section{Information Bottleneck}
Effective representation of data is key to generalizable learning. Characterizing what makes such representation good and how it emerges is crucial to understanding the success of modern machine learning. The information bottleneck (IB) method---an information-theoretic formulation for representation learning~\cite{tishby:99}---has proved a particularly useful conceptual framework for this question, and has led to a deeper understanding of representation learning in both supervised and self-supervised learning~\cite{Achille:18,Achille:18a,tian:20}. Investigating this abstraction of representation learning has the potential to yield new insights that are applicable to learning problems. 

Quantifying the goodness of a representation requires the knowledge of what is to be learned from data. Information bottleneck theory exploits the fact that, in many settings, we can define relevant information through an additional variable; for example, it could be the label of each image in a classification task. This notion of relevance allows for a precise definition of optimality{\textemdash}an IB optimal representation $T$ is maximally predictive of the relevance variable $Y$ while minimizing the number of bits extracted from the data $X$. The IB method formulates this principle as an optimization problem~\cite{tishby:99},
\begin{equation}\label{eq:ib_loss}
    \min\nolimits_{Q_{T|X}}\; I(T;X)-\beta I(T;Y).
\end{equation}
Here the optimization is over the encoders $Q_{T|X}$ which provide a (stochastic) mapping from $X$ to $T$. Maximizing the mutual information $I(T;Y)$ [second term in Eq~\eqref{eq:ib_loss}] encourages a representation $T$ to encode more relevant information while minimizing $I(T;X)$ [first term in Eq~\eqref{eq:ib_loss}] discourages it from encoding irrelevant bits. The parameter $\beta>0$ controls the fundamental tradeoff between the two information terms.

The IB method has proved successful in a number of applications, including 
neural coding~\cite{plamer:15,wang:21}, 
statistical physics~\cite{still:12,gordon:21,kline:22}, 
clustering~\cite{Strouse:19}, 
deep learning~\cite{alemi:17,Achille:18,Achille:18a},
reinforcement learning~\cite{goyal:19} and 
learning theory~\cite{bialek:01,shamir:10,bialek:20,ngampruetikorn:22}.
However the nonlinear nature of the IB problem makes it computationally costly. Although scalable learning methods based on the IB principle are possible thanks to variational bounds of mutual information~\cite{alemi:17,chalk:16,poole:19}, the choice of such bounds as well as specific details on their implementations can introduce strong inductive bias that competes with the original objective~\cite{tschannen:20}.

While large-scale applications of the IB method in its exact form are generally intractable, special cases exist.
For example, the limit of low information{\textemdash}i.e., when both terms in Eq~\eqref{eq:ib_loss} are small{\textemdash}can be described by a perturbation theory, which provides a recipe for identifying a representation that yields maximum relevant information per extracted bit~\cite{wu:19,ngampruetikorn:21}.
But perhaps the most important special case is when the source $X$ and the target $Y$ are Gaussian correlated random variables. In this case, an exact \emph{analytical} solution exists~\cite{chechik:05}. 

Although originally formulated with Shannon mutual information, the fundamental tradeoff in the IB method applies more generally: the IB optimization [Eq~\eqref{eq:ib_loss}] remains well-defined when the information terms are replaced by appropriate mutual dependence measures. 
In this work, we consider generalized IB problems based on two important correlation measures. 
The first is a parametric generalization of Shannon information, based on \renyi\ divergence~\cite{renyi:61}. \renyi-based generalizations of mutual information and entropy are central in quantifying quantum entanglement~\cite{horodecki:09,eisert:10} and have proved a powerful tool in Monte-Carlo simulations~\cite{hastings:10,singh:11,herdman:17} as well as in experiments~\cite{islam:15,bergschneider:19,brydges:19}.
The second mutual dependence measure we consider is based on Jeffreys divergence~\cite{jeffreys:46}. The resulting Jeffreys information is (up to a constant prefactor) equal to the generalization gap of a broad family of learning algorithms, known as Gibbs algorithms~\cite{aminian:21}.
We derive an analytical IB solution for the case in which $X$ and $Y$ are Gaussian correlated, generalizing the result of Ref~\cite{chechik:05} to a class of information-theoretic mutual dependence measures which includes Shannon information as a limiting case.
We show that, for both \renyi\ and Jeffreys cases, an optimal encoder can be constructed from the eigenmodes of the normalized regression matrix \smash{$\Sigma_{X|Y}\Sigma_X^{-1}$}. 
Our solution reveals a series of phase transitions, similar to those observed in the Gaussian IB method~\cite{chechik:05}.
In both \renyi\ and Jeffreys cases, we find that although the optimal encoders depend on information measures, the phase transitions occur at the critical tradeoff parameters \smash{$\beta_c^{(i)}$} that coincide with that of the Shannon case, independent of the order of \renyi\ information. 
%

\section{Divergence-based Correlation measure}
When two random variables $X$ and $Y$ are uncorrelated, their joint distribution $P_{XY}$ is equal to the product of their marginals $P_X$ and $P_Y$. As a result, we can quantify the mutual dependence between $X$ and $Y$ by the difference between $P_{XY}$ and $P_X\otimes P_Y$,
\begin{equation}
    \Omega(X;Y)\equiv \mathcal{D}(P_{XY}\parallel P_X\otimes P_Y)\ge0.
\end{equation}
Here $\mathcal{D}(P\,\|\,Q)$ denotes a statistical divergence which, by definition, is nonnegative and vanishes if and only if $P=Q$. When defined with the Kullback–Leibler (KL) divergence, the above measure becomes Shannon information,  $I(X;Y)=\KL(P_{XY}\,\|\,P_X\otimes P_Y)$.

\section{\renyi\ \textit{q}--information}

We consider a correlation measure, based on \renyi\ divergence~\cite{renyi:61}. More precisely, we define \emph{\renyi\ $q$-information}~as 
\begin{equation}\label{eq:renyi_info}
I_q(X;Y) \equiv \mathcal{R}_q(P_{XY}\parallel P_X\otimes P_Y),
\end{equation}
where $\mathcal{R}_q$ denotes \renyi\ divergence of order $q$,
\begin{equation}\label{eq:renyi_divergence}
\mathcal{R}_q(P\parallel Q)
=
\frac{1}{q-1}\ln \int\!\!dQ\, \left(\frac{dP}{dQ}\right)^q
\end{equation}
for $q\in(0,1)\cup(1,\infty)$. This definition extends to $q=0,1$ and $\infty$ via continuity in $q$. In particular, $\mathcal{R}_1(P\|Q)=\KL(P\|Q)$~\cite[Thm 5]{vanerven:14}, and as a result $I_1(X;Y)=I(X;Y)$. \renyi\ divergences, and thus $q$-information, satisfy the data processing inequality since they have a strictly increasing relationship with an $f$-divergence [with $f(t)=(t^q-1)/(q-1)$] which exhibits this property, see, e.g., Ref~\cite{liese:06}.

\subsection{Gaussian variables}

For Gaussian correlated variables $X$ and $Y$, the $q$-information is given by~(see Appendix~\ref{appx:renyi} for derivation)
\begin{equation}\label{eq:I_q_gauss}
I_q(X;Y) = \frac{-1}{2 \bar q}
\ln\frac{
|\Sigma_{X|Y}\Sigma_X^{-1}|^{\bar q}
}{
|I-\bar{q}^2(I-\Sigma_{X|Y}\Sigma_X^{-1})|
}
\;\;\text{with}\;\;
\bar q=1-q,
\end{equation}
where $I$ denotes the identity matrix in compatible dimensions. We see that this information depends on the covariance matrices only through the normalized regression matrix $\Sigma_{X|Y}\Sigma_X^{-1}$. We note also that this information can diverge when $q>2$ since the eigenvalues of $\Sigma_{X|Y}\Sigma_X^{-1}$ range from zero to one~\cite[Lemma~B.1]{chechik:05}. 
It is easy to verify that Shannon information corresponds to the limit $q\to1$, 
\begin{align}
I(X;Y) =\lim_{q\to1} I_q(X;Y) = 
-\frac{1}{2}\ln|\Sigma_{X|Y}\Sigma_X^{-1}|.
\end{align}
In addition, we note that for Gaussian variables $I_2(X;Y)=2I(X;Y)$ and $I_q(X;Y)$ increases with $q$ from zero at $q=0$.

Note that alternative definitions of \renyi\ mutual information exist. In physics literature, a frequently used definition is $I_q(X;Y)=S_q(X)+S_q(Y)-S_q(X,Y)$ where $S_q(X)=(1-q)^{-1}\ln\int dx\, p_X(x)^q$ is \renyi\ (differential) entropy of order $q$. However, for Gaussian variables, this definition leads to \renyi\ information that is equal to Shannon information regardless of $q$; the resulting \renyi\ IB problem is therefore identical to the original IB problem.

\begin{figure}
\centering
\includegraphics{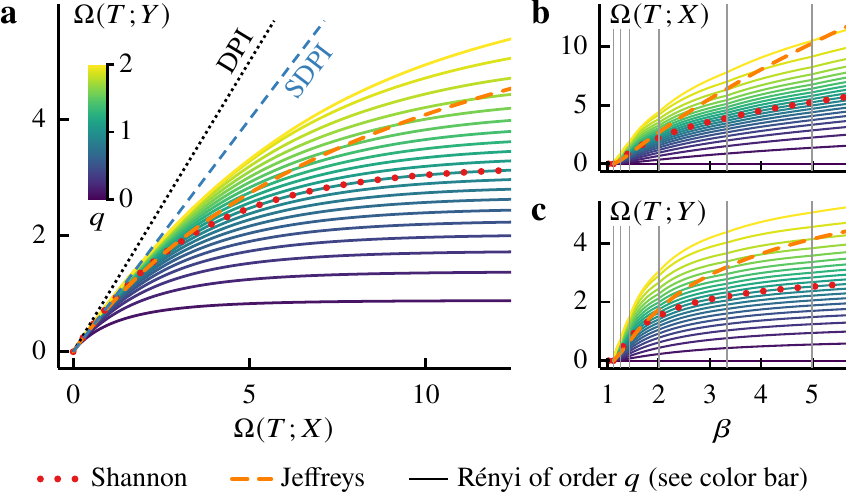}
\caption{\label{fig:ib}%
\textbf{a} The optimal frontiers of the generalized IB methods based on generalized correlation measures $\Omega(A;B)$, including Shannon ($\Omega\!=\!I$), Jeffreys ($\Omega\!=\!J$) and \renyi\ ($\Omega\!=\!I_q$) informations (see legend). For the \renyi\ case, we depict the results for a range of \renyi\ orders $q$ (see color bar).
We emphasize that while Shannon information is equivalent to \renyi\ information of order one ($q\!=\!1$), Jeffreys information is not a special case of \renyi\ information.
The relevant information $\Omega(T;Y)$ is bounded by the data processing inequality (DPI, black dotted line), $\Omega(T;Y)\!\le\!\Omega(T;X)$. We also depict the tight, data-dependent version of DPI, the strong data processing inequality (SDPI, blue dahsed line), $\Omega(T;Y)\!\le\!(1-\lambda_\text{min})\Omega(T;X)$, where $\lambda_\text{min}$ is the smallest eigenvalue of the normalized regression matrix $\Sigma_{X|Y}\Sigma_X^{-1}$.
Note that the DPI and SDPI are the same for all information measures shown.
\textbf{b-c} The extracted and relevant bits, $\Omega(T;X)$ and $\Omega(T;Y)$ respectively, increase with the tradeoff parameter $\beta$ and vanish below the critical value $\beta_c\!=\!1/(1-\lambda_\text{min})$. The vertical lines mark the location of the critical tradeoff parameters [Eqs~\eqref{eq:beta_c}\,\&\,\eqref{eq:beta_c_JIB}].
Here the eigenvalues of $\Sigma_{X|Y}\Sigma_X^{-1}$ are $\lambda_i=0.1, 0.2, 0.3, 0.5, 0.7, 0.8$.%
}
\end{figure}

\section{\renyi\ Information Bottleneck for Gaussian variables}
Replacing the mutual information in the original IB objective [Eq~\eqref{eq:ib_loss}] with $q$-information yields 
\begin{equation}\label{eq:q_IB_loss}
\mathcal{L}_{q}[Q_{T|X}] = I_q(T;X) - \beta I_{q}(T;Y)
\end{equation}
where $X$ denotes the source data, $Y$ the target variable and $T$ the representation of $X$. The loss function varies with the encoder $Q_{T|X}$ which provides a stochastic mapping from $X$ to $T$. In general, the $q$-information terms need not be of the same order but the data processing inequality $I_{q}(T;X) \ge I_{q'}(T;Y)$ is guaranteed only when $q=q'$.

We specialize to the case where $X$ and $Y$ are Gaussian correlated and consider a family of noisy linear encoders, 
\begin{equation}
T=AX+\xi \quad\text{with}\quad \xi\sim N(0,\Sigma_\xi).
\end{equation}
Since \renyi\ divergences are invariant under an invertible transformation of random variables [see Eq~\eqref{eq:renyi_divergence}], we can transform $T$ such that $\Sigma_\xi$ becomes the identity matrix without changing the information content. In the following analysis, we set $\Sigma_\xi=I$ without loss of generality. That is, the encoder becomes a Gaussian channel, parametrized only by the matrix $A$, 
\begin{equation}\label{eq:linear_encoder}
T\mid X \sim N(AX,I).
\end{equation}
To compute the information in Eq~\eqref{eq:q_IB_loss}, we first marginalize out $X$ from the above equation,  yielding
\begin{align}
T&\sim N(A\mu_X,I+A\Sigma_XA\tran)
\\
T\mid Y &\sim N(A\mu_{X|Y},I+A\Sigma_{X|Y}A\tran),
\end{align}
where we use $X\sim N(\mu_X,\Sigma_X)$ and $X\,|\,Y\sim N(\mu_{X|Y},\Sigma_{X|Y})$.
Substituting the covariance matrices in the above equations into Eq~\eqref{eq:I_q_gauss} results in 
\begin{align}\label{eq:I_q_tx}
I_q(T;X)
&=
\frac{-1}{2\bar q}
\ln
\frac{
    |I+A\Sigma_XA\tran|^q
}{
    |I+(1-\bar{q}^2)A\Sigma_XA\tran|
}
\\\label{eq:I_q_ty}
I_q(T;Y)
&=
\frac{-1}{2\bar q}
\ln
\frac{
    |I+A\Sigma_{X|Y}A\tran|^{\bar{q}}|I+A\Sigma_XA\tran|^{q}
}{
    |I+A[I-\bar{q}^2(I-\Sigma_{X|Y}\Sigma_X^{-1})]\Sigma_XA\tran|
}.
\end{align}

Following the analysis of Ref~\cite{chechik:05}, we define the \emph{mixing matrix} $W$ such that
\begin{equation}\label{eq:W_def}
A = WV,
\end{equation}
where $V$ is a matrix of left (row) eigenvectors of the normalized regression matrix $\Sigma_{X|Y}\Sigma_X^{-1}$, i.e.,
\begin{equation}\label{eq:v_def}
V\Sigma_{X|Y}\Sigma_X^{-1} = \Lambda V
\quad\text{with}\quad
\Lambda=\diag(\lambda_1,\lambda_2,\cdots).
\end{equation}
We note that $V\Sigma_X^{1/2}$ is orthogonal and thus $V\Sigma_XV\tran$ is a diagonal matrix~\cite[Lemma~B.1]{chechik:05}, i.e.,
\begin{equation}\label{eq:R_def}
V\Sigma_XV\tran = R
\quad\text{with}\quad
R = \diag(r_1,r_2,\cdots).
\end{equation}
Writing Eqs~(\ref{eq:I_q_tx}-\ref{eq:I_q_ty}) in terms of $W$, $\Lambda$ and $R$ leads to 
\begin{align}\label{eq:I_q_tx_w}
I_q(T;X)
&=
\frac{-1}{2\bar q}\ln
\frac{
    |I+WRW\tran|^q
}{
    |I+(1-\bar{q}^2)WRW\tran|
}
\\\label{eq:I_q_ty_w}
I_q(T;Y)
&=
\frac{-1}{2\bar q}\ln
\frac{
    |I+W\Lambda RW\tran|^{\bar{q}}|I+WRW\tran|^{q}
}{
    |I+W[I-\bar{q}^2(I-\Lambda)]RW\tran|
}.
\end{align}

Substituting Eqs~(\ref{eq:I_q_tx_w}-\ref{eq:I_q_ty_w}) into Eq~\eqref{eq:q_IB_loss} and setting its first order derivative with respect to the mixing matrix $W$ to zero yields the first order condition
\begin{equation}
\label{eq:first-order_cond}
\begin{aligned}
&\frac{q}{I+RW\tran W}
-
\frac{1-\bar{q}^2}{I+(1-\bar{q}^2)RW\tran W}
\\&\quad
=
\beta
\bigg(
\frac{q}{I+RW\tran W}
+
\frac{\bar{q}}{I+\Lambda RW\tran W}\Lambda 
\\&\quad\qquad
-
\frac{1}{I+[I-\bar{q}^2(I-\Lambda)]RW\tran W}
[I-\bar{q}^2(I-\Lambda)]
\bigg).
\!
\end{aligned}
\end{equation}
In deriving the above, we use the identity $d\ln|I+WCW\tran|/dW=2W(I+CW\tran W)^{-1}C$ for any compatible square matrix $C$ and assume that $R$ and $W$ are invertible. 

We seek a solution of the form
\begin{equation}\label{eq:u_def}
RW\tran W =\diag(r_1w_1^2,r_2w_2^2,\cdots) \equiv \diag(u_1,u_2,\cdots).
\end{equation}
Substituting this ansatz into Eq~\eqref{eq:first-order_cond} results in
\begin{equation}\label{eq:first-order_beta}
\frac{1}{\beta}
=
g_q(u_i,\lambda_i)
=  \frac{1-\lambda_i}{1+u_i\lambda_i}
\frac{
1+\frac{\bar{q}(1+\bar{q})(1-\lambda_i)u_i}{1+(1-\bar{q}^2(1-\lambda_i))u_i}
}{
1+\frac{\bar{q}(1+\bar{q})u_i}{1+(1-\bar{q}^2)u_i}
}.
\end{equation}
We see that the contributions from the eigenmodes of $\Sigma_{X|Y}\Sigma_X^{-1}$ decouple from one another and the reduced mixing weight, $u_i=r_iw_i^2$, for each mode depends only on the eigenvalue of that mode $\lambda_i$, the IB tradeoff parameter $\beta$ and the order of \renyi\ information $q$. 
For $\lambda\in(0,1)$ and $q\in[0,2]$, the function $g_q(u,\lambda)$ is strictly decreasing in $u$ for $u\ge0$ and approaches zero as $u\to\infty$ (see Appendix~\ref{fig:g_func}). As a result, Eq~\eqref{eq:first-order_beta} has exactly one positive solution $u_i>0$ when $1/\beta < g_q(0,\lambda_i)$. That is, the eigenmode with eigenvalue $\lambda_i$ contributes to the \renyi\ IB encoder only when $\beta$ exceeds the critical value
\begin{equation}\label{eq:beta_c}
\beta_c^{(i)} = \frac{1}{g_q(0,\lambda_i)} = \frac{1}{1-\lambda_i}.
\end{equation}
Note that $\beta_c^{(i)}$ does not depend on $q$. 
To obtain $u_i$, we can either directly solve Eq~\eqref{eq:first-order_beta} or use the analytical formula for the roots of the equivalent cubic equation (omitting the eigenmode indices)
\begin{equation}
0 = a u^3 + b u^2 + c u + d,
\end{equation}
where the coefficients are given by
\begin{align*}
a&=\lambda(1+\bar{q})(1-(1-\lambda)\bar{q}^2)
\\
b&=
\lambda(2+2\bar{q}+\lambda\bar{q}^2)
+d(1+\bar{q})(1-\lambda \bar{q}-(1-\lambda)\bar{q}^2)
\\
c&=\lambda(1+\bar{q}+\bar{q}^2) + d(2+(1-\lambda)\bar{q}-\bar{q}^2)
\\
d&=1-\beta(1-\lambda).
\end{align*}

Although the above calculation does not uniquely determine the mixing matrix $W$, we can obtain a valid IB encoder by taking $W=\diag(w_1,w_2,\cdots)$ where $w_i=\sqrt{u_i/r_i}$ since the \renyi-IB loss depends on $W$ only through the diagonal entries of $W\tran W$. To see this, we substitute Eq~\eqref{eq:u_def} into Eqs~(\ref{eq:I_q_tx_w}-\ref{eq:I_q_ty_w}) and write down
\begin{align}\label{eq:I_q_tx_u}
I_q(T;X)
&=
\frac{-1}{2\bar q}
\sum_i^{\;\;\beta>\beta_c^{(i)}}
\ln
\frac{
    (1+u_i)^q
}{
    1+(1-\bar{q}^2)u_i
}
\\\label{eq:I_q_ty_u}
I_q(T;Y)
&=
\frac{-1}{2\bar q}
\sum_i^{\;\;\beta>\beta_c^{(i)}}
\ln
\frac{
    (1+\lambda_iu_i)^{\bar{q}}(1+u_i)^{q}
}{
    1+[1-\bar{q}^2(1-\lambda_i)]u_i
},
\end{align}
where the summations are restricted to the eigenmodes that contribute the IB encoder, i.e., those with $u_i>0$. We depict the optimal frontiers of \renyi\ IB in Fig~\ref{fig:ib}.

To complete our analysis of \renyi\ IB, we note that the analytical solution of Ref~\cite{chechik:05} is a limiting case of our results. In the limit $q\to1$, Eq~\eqref{eq:first-order_beta} reads
\begin{equation}
\frac{1}{\beta}=g_{q=1}(u_i,\lambda_i)
=  \frac{1-\lambda_i}{1+u_i\lambda_i}
\implies
u_i^{(q=1)}=\frac{\beta(1-\lambda_i)-1}{\lambda_i}.
\end{equation}
Recalling that $u_i=r_iw_i^2$, we see immediately that this solution is identical to that in Ref~\cite[Lemma~4.1]{chechik:05}.

\begin{figure}
\centering
\includegraphics{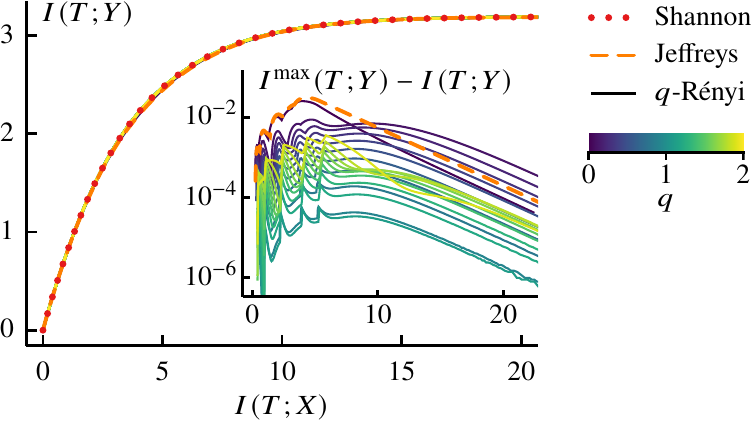}
\caption{\label{fig:shannon}%
Solving \renyi\ and Jeffreys IB problems yields representations that are close to Shannon IB optimal. Plotted on the Shannon information plane, the solutions to Shannon (dotted), Jeffreys (dashed) and \renyi\ (solid) IB problems are nearly indistinguishable. For the \renyi\ case, we depict the results for a range of \renyi\ orders $q$ (see color bar). 
\textit{Inset:} We depict the gap between the maximum achievable and encoded relevant Shannon informations, $I^\text{max}(T;Y)$ and $I(T;Y)$ respectively, as a function of the extracted Shannon information $I(T;X)$. This gap vanishes in the low and high-information limits, $I(T;X)\!\to\!0$ and $I(T;X)\!\to\!\infty$. The satellite peaks result from the fact that the solutions to Shannon, Jeffreys and \renyi\ IB problems go through structural transitions at different values of $I(T;X)$ even though these transitions occur at the same set of critical tradeoff parameters.
Here the eigenvalues of $\Sigma_{X|Y}\Sigma_X^{-1}$ are $\lambda_i=0.1, 0.2, 0.3, 0.5, 0.7, 0.8$.
}
\end{figure}

\section{Jeffreys Information Bottleneck for Gaussian variables}

The technique in the previous section applies also to the IB problems, based on other statistical divergences. In this section, we consider Jeffreys IB, defined by the loss function
\begin{equation}\label{eq:J_IB_loss}
\mathcal{L}_\mathrm{J}[Q_{T|X}] = J(T;X) - \beta J(T;Y).
\end{equation}
Here $J(X;Y)$ denotes Jeffreys information which is a mutual dependence measure, defined by
\begin{equation}
J(X;Y)\equiv \D_\mathrm{J}(P_{XY}\parallel P_X\otimes P_Y),
\end{equation}
where $\D_\mathrm{J}$ is Jeffreys divergence~\cite{jeffreys:46},
\begin{equation}
\D_\mathrm{J}(P\parallel Q)
=
\frac{1}{2}[\KL(P\parallel Q)+\KL(Q\parallel P)].
\end{equation}

For Gaussian correlated random variables, Jeffreys information takes a simple form~(see Appendix~\ref{appx:jeffreys_gauss} for derivation)
\begin{equation}
J(X;Y) 
= \frac{1}{2}\tr \left(\Sigma_X\Sigma_{X|Y}^{-1}-I\right).
\end{equation}
Using the linear encoder from Eq~\eqref{eq:linear_encoder}, the information terms in Eq~\eqref{eq:J_IB_loss} read
\begin{align}\label{eq:J_tx}
J(T;X) 
&= \frac{1}{2}\tr \left(WRW\tran\right)
\\\label{eq:J_ty}
J(T;Y) 
&= \frac{1}{2}\tr \left((I+WRW\tran)\frac{1}{I+W\Lambda RW\tran}-I\right).
\end{align}
where $W$, $\Lambda$ and $R$ are defined in Eqs~(\ref{eq:W_def}-\ref{eq:R_def}).

To solve the IB optimization, we differentiate of the loss function with respect to the mixing matrix $W$ and set the resulting derivative to zero, yielding  
\begin{equation}
I
= 
\beta \frac{1}{I+\Lambda RW\tran W}
\left(
I
-
(I+RW\tran W)
\frac{1}{I+\Lambda RW\tran W}
\Lambda
\right),
\end{equation}
where we use the identities $\partial A^{-1}/\partial a = -A^{-1} (\partial A/\partial a) A^{-1}$ and $\partial\tr(C ABA\tran)/\partial A=2CAB$ for symmetric matrices $B$ and $C$. 
We see that this equation is solvable by taking $W\tran W$ to be diagonal. Substituting Eq~\eqref{eq:u_def} into the above equation and solving for $u_i$ gives
\begin{equation}\label{eq:JIB_u}
u_i
= 
\frac{\sqrt{\beta(1-\lambda_i)}-1}{\lambda_i}.
\end{equation}
Since $u_i=r_iw_i^2\ge0$, we see that this solution is valid only when the tradeoff parameter $\beta$ in greater than the critical value 
\begin{equation}\label{eq:beta_c_JIB}
\beta_c^{(i)}
= 
\frac{1}{1-\lambda_i}.
\end{equation}
We note that this critical value is identical to that of \renyi\ IB [Eq~\eqref{eq:beta_c}]. 
Finally substituting Eq~\eqref{eq:JIB_u} into Eqs~(\ref{eq:J_tx}-\ref{eq:J_ty}) leads to
\begin{align}
J(T;X) &=
\frac{1}{2}\sum_i^{\;\;\beta>\beta_c^{(i)}}
\frac{\sqrt{\beta(1-\lambda_i)}-1}{\lambda_i}
\\
J(T;Y) &=
\frac{1}{2}\sum_i^{\;\;\beta>\beta_c^{(i)}}
\frac{1-\lambda_i}{\lambda_i}
\frac{\sqrt{\beta(1-\lambda_i)}-1}{\sqrt{\beta(1-\lambda_i)}}.
\end{align}
where the summations are limited to the modes that contribute to the encoder, i.e., those with $\beta_c^{(i)}<\beta$. 
In Fig~\ref{fig:ib}, we depict an example of the Jeffreys IB optimal frontier, computed from the above equations. We emphasize that while Shannon information is equivalent to \renyi\ information with $q=1$, Jeffreys information is not a special case of \renyi\ information.

\section{Discussion \& Conclusion}

In Fig~\ref{fig:shannon}, we depict the solutions to the original, \renyi\ and Jeffreys IB problems on the Shannon information plane. We see that these solutions are very close to the optimal frontier, characterized by the Shannon IB solutions. This result suggests that formulating and solving an IB problem, defined with alternative correlation measures other than Shannon information, could offer a strategy for obtaining an approximate solution to the original IB problem.
To better illustrate the differences between the solutions to the original, \renyi\ and Jeffreys IB problems, the inset shows how much less relevant Shannon information the optimal representations of \renyi\ and Jeffreys IB encode, compared to the Shannon IB optimal representation. We see that the differences are maximum at intermediate information and vanish in the low and high-information limits. In addition, the Shannon information gaps exhibit satellite peaks, resulting from structural the transition of the IB solutions. We note that although these transitions occur at the same critical tradeoff parameters [Eqs~\eqref{eq:beta_c}\,\&\,\eqref{eq:beta_c_JIB}], they generally correspond to different values of extracted Shannon information.

To sum up, we consider generalized IB problems in which the mutual information is replaced by mutual dependence measures, based on \renyi\ and Jeffreys divergences. We obtain exact analytical solutions for the case of Gaussian correlated random variables, generalizing the results of Ref~\cite{chechik:05}. We show that the fundamental IB tradeoff between relevance and compression holds also for correlation measures other than Shannon information. Our analyses reveal structural transitions in the optimal representations, similar to that in the original IB method \cite{chechik:05}. Interestingly the critical tradeoff parameters are the same for original, \renyi\ and Jeffreys IB problems, even though the solutions are distinct.

We anticipate that our work will find application in physics of correlated components which relies on \renyi-generalization of entropy and information to quantify entanglement. In addition, our characterization of Jeffreys IB could have implications for understanding the generalization properties of Gibbs learning algorithms of which the generalization gap is proportional to Jeffreys information between fitted models and training data. Finally we note that the conditional IB problem, in which the compression term $I(T;X)$ is replaced by $I(T;X\,|\,Y)$, becomes non-trivial for generalized information measures since the chain rule does not hold for \renyi\ and Jeffreys information---that is, given the Markov constraint $T$--$X$--$Y$, we have $I(T;X\,|\,Y)=I(T;X)-I(T;Y)$ for Shannon information, but in general, $\Omega(T;X\,|\,Y)\ne\Omega(T;X)-\Omega(T;Y)$. The logical steps in our analyses are readily generalizable to conditional IB problems.

\begin{acknowledgments}
This work was supported in part by the National Science Foundation, through the Center for the Physics of Biological Function (PHY-1734030), the Simons Foundation and the Sloan Foundation.
\end{acknowledgments}

\appendix

\section{\renyi\ information for Gaussian variables\label{appx:renyi}}
In this appendix, we derive \renyi\ mutual information for Gaussian correlated variables. Using the definition from Eqs~(\ref{eq:renyi_info}-\ref{eq:renyi_divergence}), we write down \renyi\ mutual information for continuous random variables,
\begin{equation}\label{eq:renyi_info_continuous}
I_q(X;Y) =
\frac{1}{q-1}\ln \int\! dx dy\, p_X(x)p_Y(y)
\left(\frac{p_{XY}(x,y)}{p_X(x)p_Y(y)}\right)^q.
\end{equation}
where $p_X$, $p_Y$ and $p_{XY}$ denote the probability density functions of $X$, $Y$ and $(X,Y)$, respectively. We consider Gaussian correlated random variables 
\begin{equation}\label{eq:xy_gauss}
\genfrac{[}{]}{0pt}{0}{X}{Y}
\sim
N(\mu,\Sigma)
\;\;\text{with}\;\;
\mu=\genfrac{[}{]}{0pt}{0}{\mu_X}{\mu_Y}
\;\;\text{and}\;\;
\Sigma=\left[
\begin{matrix}
\Sigma_X&\Sigma_{XY}\\\Sigma_{YX}&\Sigma_{Y}
\end{matrix}
\right].
\end{equation}
In this case, the joint probability density is given by
\begin{equation}\label{eq:P_XY}
p_{XY}(x,y) = 
\frac{\exp
\left\{
- 
([\begin{smallmatrix}x\\y\end{smallmatrix}]-\mu)\tran
\Sigma^{-1}
([\begin{smallmatrix}x\\y\end{smallmatrix}]-\mu)
\right\}
}{
|2\pi\Sigma|^{1/2}}
\end{equation}
The product of the marginal distributions is equal to a joint distribution but with $\Sigma_{XY}$ and $\Sigma_{YX}$ set to zero, i.e.,
\begin{equation}\label{eq:P_XP_Y}
p_X(x)p_Y(y) = \frac{
\exp
\left\{
- 
([\begin{smallmatrix}x\\y\end{smallmatrix}]-\mu)\tran
\bar\Sigma^{-1}
([\begin{smallmatrix}x\\y\end{smallmatrix}]-\mu)
\right\}
}{
|2\pi\bar\Sigma|^{1/2}
}
\end{equation}
where $\bar\Sigma=\left[
\begin{smallmatrix}
\Sigma_X&\cdot\\\cdot&\Sigma_{Y}
\end{smallmatrix}\right]$. 
Substituting the above densities into Eq~\eqref{eq:renyi_info_continuous} and performing the resulting Gaussian integration over $x$ and $y$ gives
\begin{equation}\label{eq:renyi_info_gauss_0}
I_q(X;Y) =
\frac{1}{q-1}\ln 
\frac{
|q\Sigma^{-1}+(1-q)\bar\Sigma^{-1}|^{-1/2}
}{
|\Sigma|^{q/2}
|\bar\Sigma|^{(1-q)/2}
}.
\end{equation}
The determinants of the covariance matrices are given by
\begin{equation}\label{eq:det_Sigma}
|\Sigma|=|\Sigma_Y|\times|\Sigma_{X|Y}|
\quad\text{and}\quad
|\bar\Sigma|=|\Sigma_Y|\times|\Sigma_{X}|,
\end{equation}
where $\Sigma_{X|Y}=\Sigma_X-\Sigma_{XY}\Sigma_Y^{-1}\Sigma_{YX}$ and we use the identity
\begin{equation}\label{eq:det_identity}
\left|\begin{matrix}A&B\\C&D\end{matrix}\right|
=
|D|\times|A-BD^{-1}C|.
\end{equation}
We now consider the numerator in Eq~\eqref{eq:renyi_info_gauss_0},
\begin{align}
\left|q\Sigma^{-1}+(1-q)\bar\Sigma^{-1}\right|
&=
\left|\Sigma^{-1}
\left(q\bar\Sigma+(1-q)\Sigma\right)
\bar\Sigma^{-1}\right|
\nonumber\\
&=
\frac{1}{|\Sigma|\times|\bar\Sigma|}
\left|
\begin{matrix}
\Sigma_X&(1-q)\Sigma_{XY}
\nonumber\\
(1-q)\Sigma_{YX}&\Sigma_{Y}
\end{matrix}
\right|
\\
&=\frac{
\left|
I-(1-q)^2(I-\Sigma_{X|Y}\Sigma_X^{-1})
\right|
}{
|\Sigma_Y|\times|\Sigma_{X|Y}|
},
\end{align}
where the last equality follows from Eqs~(\ref{eq:det_Sigma}-\ref{eq:det_identity}).
Finally we write down the \renyi\ information for Gaussian variables
\begin{equation}
I_q(X;Y) =
\frac{1/2}{q-1}\ln 
\frac{
|\Sigma_{X|Y}\Sigma_{X}^{-1}|^{1-q}
}{
\left|
I-(1-q)^2(I-\Sigma_{X|Y}\Sigma_X^{-1})
\right|
}.
\end{equation}
This expression is identical to Eq~\eqref{eq:I_q_gauss} (with $\bar q=1-q$).

\section{Jeffreys information for Gaussian variables\label{appx:jeffreys_gauss}}
The Jeffreys information is defined via
\begin{equation}
J(X;Y)\equiv \D_\mathrm{J}(P_{XY}\parallel P_X\otimes P_Y),
\end{equation}
where $\D_\mathrm{J}$ is Jeffreys divergence~\cite{jeffreys:46},
\begin{equation}\label{eq:DJ_defo}
\D_\mathrm{J}(P\parallel Q)
=
\frac{1}{2}[\KL(P\parallel Q)+\KL(Q\parallel P)].
\end{equation}
For Gaussian correlated $X$ and $Y$, the Jeffreys information follows immediately from the KL divergence between two multivariate Gaussian distributions
\begin{align}
&\KL(N(\mu_0,\Sigma_0)\parallel N(\mu_1,\Sigma_1))
=
\frac{1}{2}\bigg(
\tr(\Sigma_1^{-1}\Sigma_0 - I )
\nonumber\\&\hspace{2cm}
+ (\mu_1-\mu_0)\tran \Sigma_1^{-1} (\mu_1-\mu_0)
+ \ln \frac{|\Sigma_1|}{|\Sigma_0|}
\bigg).
\end{align}
For $X$ and $Y$ described by Eq~\eqref{eq:xy_gauss}, we have $P_{XY}=N(\mu,\Sigma)$ and $P_{X}\otimes P_Y=N(\mu,\bar\Sigma)$, where $\Sigma=\left[
\begin{smallmatrix}
\Sigma_X&\Sigma_{XY}\\\Sigma_{YX}&\Sigma_{Y}
\end{smallmatrix}\right]$ 
and 
$\bar\Sigma=\left[
\begin{smallmatrix}
\Sigma_X&\cdot\\\cdot&\Sigma_{Y}
\end{smallmatrix}\right]$. 
As a result, we have  
\begin{align}\label{eq:KL_forward}
\KL(P_{XY}\parallel P_X\otimes P_Y)
&=
\frac{1}{2}\left(
\tr(\bar\Sigma^{-1}\Sigma - I )
+ \ln \frac{|\bar\Sigma|}{|\Sigma|}
\right)
\\
\label{eq:KL_reverse}
\KL(P_X\otimes P_Y\parallel P_{XY})
&=
\frac{1}{2}\left(
\tr(\Sigma^{-1}\bar\Sigma - I )
+ \ln \frac{|\Sigma|}{|\bar\Sigma|}
\right).
\end{align}
We see that the logarithmic term drops out upon symmetrization [Eq~\eqref{eq:DJ_defo}]. 
Substituting 
$\bar\Sigma^{-1}=\left[
\begin{smallmatrix}
\Sigma_X^{-1}&\cdot\\\cdot&\Sigma_{Y}^{-1}
\end{smallmatrix}\right]$ and the determinant formula in Eq~\eqref{eq:det_Sigma} into Eq~\eqref{eq:KL_forward} gives
\begin{equation}\label{eq:shannon_info}
\KL(P_{XY}\parallel P_X\otimes P_Y)
=
\frac{1}{2}\ln \frac{|\bar\Sigma|}{|\Sigma|}
=
-\frac{1}{2}\ln |\Sigma_{X|Y}\Sigma_X^{-1} |
\end{equation}
which is the usual mutual information, as expected. To compute trace in Eq~\eqref{eq:KL_reverse}, we write down the inverse of the covariance matrix,
\begin{align}
\Sigma^{-1}=
\begin{pmatrix}
\Sigma_{X|Y}^{-1}&-\Sigma_{X|Y}^{-1}\Sigma_{XY}\Sigma_{Y}^{-1}
\\
-\Sigma_{Y}^{-1}\Sigma_{YX}\Sigma_{X|Y}^{-1}&\Sigma_{Y|X}^{-1}
\end{pmatrix}.
\end{align}
Therefore we have
\begin{align}
\tr(\Sigma^{-1}\bar\Sigma - I ) 
&=
\tr(\Sigma^{-1}(\bar\Sigma-\Sigma)) 
\nonumber\\
&=
\tr\left(
\left[
\begin{smallmatrix}
\Sigma_{X|Y}^{-1}&-\Sigma_{X|Y}^{-1}\Sigma_{XY}\Sigma_{Y}^{-1}
\\
-\Sigma_{Y}^{-1}\Sigma_{YX}\Sigma_{X|Y}^{-1}&\Sigma_{Y|X}^{-1}
\end{smallmatrix}
\right]
\left[
\begin{smallmatrix}
\cdot&-\Sigma_{XY}\\-\Sigma_{YX}&\cdot
\end{smallmatrix}\right]
\right)
\nonumber\\
&=
\tr(\Sigma_{X|Y}^{-1}\Sigma_{XY}\Sigma_{Y}^{-1}\Sigma_{YX})
+
\tr(\Sigma_{Y}^{-1}\Sigma_{YX}\Sigma_{X|Y}^{-1}\Sigma_{XY})
\nonumber\\
&=
2\tr(\Sigma_{XY}\Sigma_{Y}^{-1}\Sigma_{YX}\Sigma_{X|Y}^{-1})
\nonumber\\
&=
2\tr(\Sigma_X\Sigma_{X|Y}^{-1}-I),
\end{align}
where the last equality follows from the identity $\Sigma_{X|Y}=\Sigma_X-\Sigma_{XY}\Sigma_Y^{-1}\Sigma_{YX}$. 
Substituting the above result into Eq~\eqref{eq:KL_reverse} yields
\begin{equation}
\KL(P_X\otimes P_Y\parallel P_{XY})
=
\tr(\Sigma_X\Sigma_{X|Y}^{-1}-I)
+
\frac{1}{2}\ln |\Sigma_{X|Y}\Sigma_X^{-1} |.
\end{equation}
Finally eliminating the logarithmic term with Eq~\eqref{eq:shannon_info} leads to
\begin{align}
J(X;Y) 
&= 
\frac{1}{2}[\KL(P_X\otimes P_Y\parallel P_{XY})
+
\KL(P_{XY}\parallel P_X\otimes P_Y)]
\nonumber\\
&= \frac{1}{2}\tr \left(\Sigma_X\Sigma_{X|Y}^{-1}-I\right).
\end{align}

\section{Supplementary figure\label{fig:g_func}}
\renewcommand{\thefigure}{\thesection\arabic{figure}}
\setcounter{figure}{0}
\begin{figure}[h]
\centering
\includegraphics[width=\linewidth]{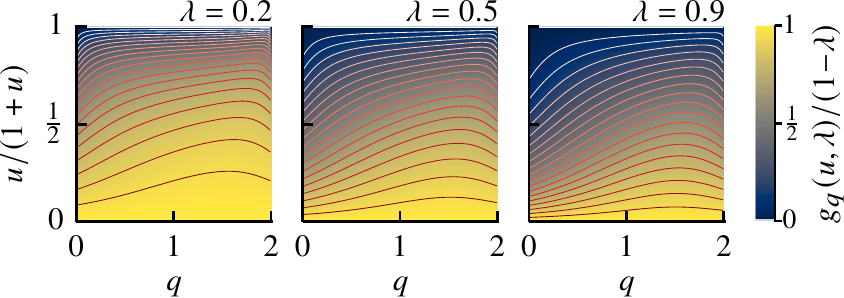}
\caption{
The function $g_q(u,\lambda)$ [Eq~\eqref{eq:first-order_beta}] decreases with $u$ from $1\!-\!\lambda$ at $u\!=\!0$ and approaches zero as $u\!\to\!\infty$. As a result, the equation $\beta^{-1}\!=\!g_q(u,\lambda)$ always has a unique positive solution when $\beta\!>\!1/(1-\lambda)$. We consider only $0\!\le\!q\!\le\!2$ since \renyi\ information for Gaussian variables can diverge for $q\!>\!2$ [see Eq~\eqref{eq:I_q_gauss}].%
}
\end{figure}
\eject


\begin{thebibliography}{35}%
\makeatletter
\providecommand \@ifxundefined [1]{%
 \@ifx{#1\undefined}
}%
\providecommand \@ifnum [1]{%
 \ifnum #1\expandafter \@firstoftwo
 \else \expandafter \@secondoftwo
 \fi
}%
\providecommand \@ifx [1]{%
 \ifx #1\expandafter \@firstoftwo
 \else \expandafter \@secondoftwo
 \fi
}%
\providecommand \natexlab [1]{#1}%
\providecommand \enquote  [1]{``#1''}%
\providecommand \bibnamefont  [1]{#1}%
\providecommand \bibfnamefont [1]{#1}%
\providecommand \citenamefont [1]{#1}%
\providecommand \href@noop [0]{\@secondoftwo}%
\providecommand \href [0]{\begingroup \@sanitize@url \@href}%
\providecommand \@href[1]{\@@startlink{#1}\@@href}%
\providecommand \@@href[1]{\endgroup#1\@@endlink}%
\providecommand \@sanitize@url [0]{\catcode `\\12\catcode `\$12\catcode
  `\&12\catcode `\#12\catcode `\^12\catcode `\_12\catcode `\%12\relax}%
\providecommand \@@startlink[1]{}%
\providecommand \@@endlink[0]{}%
\providecommand \url  [0]{\begingroup\@sanitize@url \@url }%
\providecommand \@url [1]{\endgroup\@href {#1}{\urlprefix }}%
\providecommand \urlprefix  [0]{URL }%
\providecommand \Eprint [0]{\href }%
\providecommand \doibase [0]{https://doi.org/}%
\providecommand \selectlanguage [0]{\@gobble}%
\providecommand \bibinfo  [0]{\@secondoftwo}%
\providecommand \bibfield  [0]{\@secondoftwo}%
\providecommand \translation [1]{[#1]}%
\providecommand \BibitemOpen [0]{}%
\providecommand \bibitemStop [0]{}%
\providecommand \bibitemNoStop [0]{.\EOS\space}%
\providecommand \EOS [0]{\spacefactor3000\relax}%
\providecommand \BibitemShut  [1]{\csname bibitem#1\endcsname}%
\let\auto@bib@innerbib\@empty
\bibitem [{\citenamefont {Tishby}\ \emph {et~al.}(1999)\citenamefont {Tishby},
  \citenamefont {Pereira},\ and\ \citenamefont {Bialek}}]{tishby:99}%
  \BibitemOpen
  \bibfield  {author} {\bibinfo {author} {\bibfnamefont {N.}~\bibnamefont
  {Tishby}}, \bibinfo {author} {\bibfnamefont {F.~C.~N.}\ \bibnamefont
  {Pereira}},\ and\ \bibinfo {author} {\bibfnamefont {W.}~\bibnamefont
  {Bialek}},\ }\bibfield  {title} {\bibinfo {title} {The information bottleneck
  method},\ }in\ \href {http://arxiv.org/abs/physics/0004057} {\emph {\bibinfo
  {booktitle} {37th Allerton Conference on Communication,~Control and
  Computing}}},\ \bibinfo {editor} {edited by\ \bibinfo {editor} {\bibfnamefont
  {B.}~\bibnamefont {Hajek}}\ and\ \bibinfo {editor} {\bibfnamefont {R.~S.}\
  \bibnamefont {Sreenivas}}}\ (\bibinfo  {publisher} {University of Illinois},\
  \bibinfo {year} {1999})\ pp.\ \bibinfo {pages} {368--377}\BibitemShut
  {NoStop}%
\bibitem [{\citenamefont {{Achille}}\ and\ \citenamefont
  {{Soatto}}(2018)}]{Achille:18}%
  \BibitemOpen
  \bibfield  {author} {\bibinfo {author} {\bibfnamefont {A.}~\bibnamefont
  {{Achille}}}\ and\ \bibinfo {author} {\bibfnamefont {S.}~\bibnamefont
  {{Soatto}}},\ }\bibfield  {title} {\bibinfo {title} {Information dropout:
  Learning optimal representations through noisy computation},\ }\href
  {https://doi.org/10.1109/TPAMI.2017.2784440} {\bibfield  {journal} {\bibinfo
  {journal} {IEEE Transactions on Pattern Analysis and Machine Intelligence}\
  }\textbf {\bibinfo {volume} {40}},\ \bibinfo {pages} {2897} (\bibinfo {year}
  {2018})}\BibitemShut {NoStop}%
\bibitem [{\citenamefont {Achille}\ and\ \citenamefont
  {Soatto}(2018)}]{Achille:18a}%
  \BibitemOpen
  \bibfield  {author} {\bibinfo {author} {\bibfnamefont {A.}~\bibnamefont
  {Achille}}\ and\ \bibinfo {author} {\bibfnamefont {S.}~\bibnamefont
  {Soatto}},\ }\bibfield  {title} {\bibinfo {title} {Emergence of invariance
  and disentanglement in deep representations},\ }\href
  {http://jmlr.org/papers/v19/17-646.html} {\bibfield  {journal} {\bibinfo
  {journal} {Journal of Machine Learning Research}\ }\textbf {\bibinfo {volume}
  {19}},\ \bibinfo {pages} {1} (\bibinfo {year} {2018})}\BibitemShut {NoStop}%
\bibitem [{\citenamefont {Tian}\ \emph {et~al.}(2020)\citenamefont {Tian},
  \citenamefont {Sun}, \citenamefont {Poole}, \citenamefont {Krishnan},
  \citenamefont {Schmid},\ and\ \citenamefont {Isola}}]{tian:20}%
  \BibitemOpen
  \bibfield  {author} {\bibinfo {author} {\bibfnamefont {Y.}~\bibnamefont
  {Tian}}, \bibinfo {author} {\bibfnamefont {C.}~\bibnamefont {Sun}}, \bibinfo
  {author} {\bibfnamefont {B.}~\bibnamefont {Poole}}, \bibinfo {author}
  {\bibfnamefont {D.}~\bibnamefont {Krishnan}}, \bibinfo {author}
  {\bibfnamefont {C.}~\bibnamefont {Schmid}},\ and\ \bibinfo {author}
  {\bibfnamefont {P.}~\bibnamefont {Isola}},\ }\bibfield  {title} {\bibinfo
  {title} {What makes for good views for contrastive learning?},\ }in\ \href
  {https://proceedings.neurips.cc/paper/2020/hash/4c2e5eaae9152079b9e95845750bb9ab-Abstract.html}
  {\emph {\bibinfo {booktitle} {Advances in Neural Information Processing
  Systems}}},\ Vol.~\bibinfo {volume} {33},\ \bibinfo {editor} {edited by\
  \bibinfo {editor} {\bibfnamefont {H.}~\bibnamefont {Larochelle}}, \bibinfo
  {editor} {\bibfnamefont {M.}~\bibnamefont {Ranzato}}, \bibinfo {editor}
  {\bibfnamefont {R.}~\bibnamefont {Hadsell}}, \bibinfo {editor} {\bibfnamefont
  {M.}~\bibnamefont {Balcan}},\ and\ \bibinfo {editor} {\bibfnamefont
  {H.}~\bibnamefont {Lin}}}\ (\bibinfo  {publisher} {Curran Associates, Inc.},\
  \bibinfo {year} {2020})\ pp.\ \bibinfo {pages} {6827--6839}\BibitemShut
  {NoStop}%
\bibitem [{\citenamefont {Palmer}\ \emph {et~al.}(2015)\citenamefont {Palmer},
  \citenamefont {Marre}, \citenamefont {Berry},\ and\ \citenamefont
  {Bialek}}]{plamer:15}%
  \BibitemOpen
  \bibfield  {author} {\bibinfo {author} {\bibfnamefont {S.~E.}\ \bibnamefont
  {Palmer}}, \bibinfo {author} {\bibfnamefont {O.}~\bibnamefont {Marre}},
  \bibinfo {author} {\bibfnamefont {M.~J.}\ \bibnamefont {Berry}},\ and\
  \bibinfo {author} {\bibfnamefont {W.}~\bibnamefont {Bialek}},\ }\bibfield
  {title} {\bibinfo {title} {Predictive information in a sensory population},\
  }\href {https://doi.org/10.1073/pnas.1506855112} {\bibfield  {journal}
  {\bibinfo  {journal} {Proceedings of the National Academy of Sciences}\
  }\textbf {\bibinfo {volume} {112}},\ \bibinfo {pages} {6908} (\bibinfo {year}
  {2015})}\BibitemShut {NoStop}%
\bibitem [{\citenamefont {Wang}\ \emph {et~al.}(2021)\citenamefont {Wang},
  \citenamefont {Segev}, \citenamefont {Borst},\ and\ \citenamefont
  {Palmer}}]{wang:21}%
  \BibitemOpen
  \bibfield  {author} {\bibinfo {author} {\bibfnamefont {S.}~\bibnamefont
  {Wang}}, \bibinfo {author} {\bibfnamefont {I.}~\bibnamefont {Segev}},
  \bibinfo {author} {\bibfnamefont {A.}~\bibnamefont {Borst}},\ and\ \bibinfo
  {author} {\bibfnamefont {S.}~\bibnamefont {Palmer}},\ }\bibfield  {title}
  {\bibinfo {title} {Maximally efficient prediction in the early fly visual
  system may support evasive flight maneuvers},\ }\href
  {https://doi.org/10.1371/journal.pcbi.1008965} {\bibfield  {journal}
  {\bibinfo  {journal} {PLOS Computational Biology}\ }\textbf {\bibinfo
  {volume} {17}},\ \bibinfo {pages} {e1008965} (\bibinfo {year}
  {2021})}\BibitemShut {NoStop}%
\bibitem [{\citenamefont {Still}\ \emph {et~al.}(2012)\citenamefont {Still},
  \citenamefont {Sivak}, \citenamefont {Bell},\ and\ \citenamefont
  {Crooks}}]{still:12}%
  \BibitemOpen
  \bibfield  {author} {\bibinfo {author} {\bibfnamefont {S.}~\bibnamefont
  {Still}}, \bibinfo {author} {\bibfnamefont {D.~A.}\ \bibnamefont {Sivak}},
  \bibinfo {author} {\bibfnamefont {A.~J.}\ \bibnamefont {Bell}},\ and\
  \bibinfo {author} {\bibfnamefont {G.~E.}\ \bibnamefont {Crooks}},\ }\bibfield
   {title} {\bibinfo {title} {Thermodynamics of prediction},\ }\href
  {https://doi.org/10.1103/PhysRevLett.109.120604} {\bibfield  {journal}
  {\bibinfo  {journal} {Physical Review Letters}\ }\textbf {\bibinfo {volume}
  {109}},\ \bibinfo {pages} {120604} (\bibinfo {year} {2012})}\BibitemShut
  {NoStop}%
\bibitem [{\citenamefont {Gordon}\ \emph {et~al.}(2021)\citenamefont {Gordon},
  \citenamefont {Banerjee}, \citenamefont {Koch-Janusz},\ and\ \citenamefont
  {Ringel}}]{gordon:21}%
  \BibitemOpen
  \bibfield  {author} {\bibinfo {author} {\bibfnamefont {A.}~\bibnamefont
  {Gordon}}, \bibinfo {author} {\bibfnamefont {A.}~\bibnamefont {Banerjee}},
  \bibinfo {author} {\bibfnamefont {M.}~\bibnamefont {Koch-Janusz}},\ and\
  \bibinfo {author} {\bibfnamefont {Z.}~\bibnamefont {Ringel}},\ }\bibfield
  {title} {\bibinfo {title} {Relevance in the renormalization group and in
  information theory},\ }\href {https://doi.org/10.1103/PhysRevLett.126.240601}
  {\bibfield  {journal} {\bibinfo  {journal} {Physical Review Letters}\
  }\textbf {\bibinfo {volume} {126}},\ \bibinfo {pages} {240601} (\bibinfo
  {year} {2021})}\BibitemShut {NoStop}%
\bibitem [{\citenamefont {Kline}\ and\ \citenamefont
  {Palmer}(2022)}]{kline:22}%
  \BibitemOpen
  \bibfield  {author} {\bibinfo {author} {\bibfnamefont {A.~G.}\ \bibnamefont
  {Kline}}\ and\ \bibinfo {author} {\bibfnamefont {S.~E.}\ \bibnamefont
  {Palmer}},\ }\bibfield  {title} {\bibinfo {title} {Gaussian information
  bottleneck and the non-perturbative renormalization group},\ }\href
  {https://doi.org/10.1088/1367-2630/ac395d} {\bibfield  {journal} {\bibinfo
  {journal} {New Journal of Physics}\ }\textbf {\bibinfo {volume} {24}},\
  \bibinfo {pages} {033007} (\bibinfo {year} {2022})}\BibitemShut {NoStop}%
\bibitem [{\citenamefont {Strouse}\ and\ \citenamefont
  {Schwab}(2019)}]{Strouse:19}%
  \BibitemOpen
  \bibfield  {author} {\bibinfo {author} {\bibfnamefont {D.}~\bibnamefont
  {Strouse}}\ and\ \bibinfo {author} {\bibfnamefont {D.~J.}\ \bibnamefont
  {Schwab}},\ }\bibfield  {title} {\bibinfo {title} {The information bottleneck
  and geometric clustering},\ }\href {https://doi.org/10.1162/neco_a_01136}
  {\bibfield  {journal} {\bibinfo  {journal} {Neural Computation}\ }\textbf
  {\bibinfo {volume} {31}},\ \bibinfo {pages} {596} (\bibinfo {year}
  {2019})}\BibitemShut {NoStop}%
\bibitem [{\citenamefont {Alemi}\ \emph {et~al.}(2017)\citenamefont {Alemi},
  \citenamefont {Fischer}, \citenamefont {Dillon},\ and\ \citenamefont
  {Murphy}}]{alemi:17}%
  \BibitemOpen
  \bibfield  {author} {\bibinfo {author} {\bibfnamefont {A.~A.}\ \bibnamefont
  {Alemi}}, \bibinfo {author} {\bibfnamefont {I.}~\bibnamefont {Fischer}},
  \bibinfo {author} {\bibfnamefont {J.~V.}\ \bibnamefont {Dillon}},\ and\
  \bibinfo {author} {\bibfnamefont {K.}~\bibnamefont {Murphy}},\ }\bibfield
  {title} {\bibinfo {title} {Deep variational information bottleneck},\ }in\
  \href {https://openreview.net/forum?id=HyxQzBceg} {\emph {\bibinfo
  {booktitle} {International Conference on Learning Representations}}}\
  (\bibinfo {year} {2017})\BibitemShut {NoStop}%
\bibitem [{\citenamefont {Goyal}\ \emph {et~al.}(2019)\citenamefont {Goyal},
  \citenamefont {Islam}, \citenamefont {Strouse}, \citenamefont {Ahmed},
  \citenamefont {Larochelle}, \citenamefont {Botvinick}, \citenamefont
  {Levine},\ and\ \citenamefont {Bengio}}]{goyal:19}%
  \BibitemOpen
  \bibfield  {author} {\bibinfo {author} {\bibfnamefont {A.}~\bibnamefont
  {Goyal}}, \bibinfo {author} {\bibfnamefont {R.}~\bibnamefont {Islam}},
  \bibinfo {author} {\bibfnamefont {D.}~\bibnamefont {Strouse}}, \bibinfo
  {author} {\bibfnamefont {Z.}~\bibnamefont {Ahmed}}, \bibinfo {author}
  {\bibfnamefont {H.}~\bibnamefont {Larochelle}}, \bibinfo {author}
  {\bibfnamefont {M.}~\bibnamefont {Botvinick}}, \bibinfo {author}
  {\bibfnamefont {S.}~\bibnamefont {Levine}},\ and\ \bibinfo {author}
  {\bibfnamefont {Y.}~\bibnamefont {Bengio}},\ }\bibfield  {title} {\bibinfo
  {title} {Transfer and exploration via the information bottleneck},\ }in\
  \href {https://openreview.net/forum?id=rJg8yhAqKm} {\emph {\bibinfo
  {booktitle} {International Conference on Learning Representations}}}\
  (\bibinfo {year} {2019})\BibitemShut {NoStop}%
\bibitem [{\citenamefont {Bialek}\ \emph {et~al.}(2001)\citenamefont {Bialek},
  \citenamefont {Nemenman},\ and\ \citenamefont {Tishby}}]{bialek:01}%
  \BibitemOpen
  \bibfield  {author} {\bibinfo {author} {\bibfnamefont {W.}~\bibnamefont
  {Bialek}}, \bibinfo {author} {\bibfnamefont {I.}~\bibnamefont {Nemenman}},\
  and\ \bibinfo {author} {\bibfnamefont {N.}~\bibnamefont {Tishby}},\
  }\bibfield  {title} {\bibinfo {title} {{Predictability, Complexity, and
  Learning}},\ }\href {https://doi.org/10.1162/089976601753195969} {\bibfield
  {journal} {\bibinfo  {journal} {Neural Computation}\ }\textbf {\bibinfo
  {volume} {13}},\ \bibinfo {pages} {2409} (\bibinfo {year}
  {2001})}\BibitemShut {NoStop}%
\bibitem [{\citenamefont {Shamir}\ \emph {et~al.}(2010)\citenamefont {Shamir},
  \citenamefont {Sabato},\ and\ \citenamefont {Tishby}}]{shamir:10}%
  \BibitemOpen
  \bibfield  {author} {\bibinfo {author} {\bibfnamefont {O.}~\bibnamefont
  {Shamir}}, \bibinfo {author} {\bibfnamefont {S.}~\bibnamefont {Sabato}},\
  and\ \bibinfo {author} {\bibfnamefont {N.}~\bibnamefont {Tishby}},\
  }\bibfield  {title} {\bibinfo {title} {Learning and generalization with the
  information bottleneck},\ }\href {https://doi.org/10.1016/j.tcs.2010.04.006}
  {\bibfield  {journal} {\bibinfo  {journal} {Theoretical Computer Science}\
  }\textbf {\bibinfo {volume} {411}},\ \bibinfo {pages} {2696} (\bibinfo {year}
  {2010})},\ \bibinfo {note} {{Algorithmic Learning Theory (ALT
  2008)}}\BibitemShut {NoStop}%
\bibitem [{\citenamefont {Bialek}\ \emph {et~al.}(2020)\citenamefont {Bialek},
  \citenamefont {Palmer},\ and\ \citenamefont {Schwab}}]{bialek:20}%
  \BibitemOpen
  \bibfield  {author} {\bibinfo {author} {\bibfnamefont {W.}~\bibnamefont
  {Bialek}}, \bibinfo {author} {\bibfnamefont {S.~E.}\ \bibnamefont {Palmer}},\
  and\ \bibinfo {author} {\bibfnamefont {D.~J.}\ \bibnamefont {Schwab}},\
  }\href {https://doi.org/10.48550/ARXIV.2008.12279} {\bibinfo {title} {What
  makes it possible to learn probability distributions in the natural world?}}
  (\bibinfo {year} {2020}),\ \Eprint {https://arxiv.org/abs/2008.12279}
  {arXiv:2008.12279 [cond-mat.stat-mech]} \BibitemShut {NoStop}%
\bibitem [{\citenamefont {Ngampruetikorn}\ and\ \citenamefont
  {Schwab}(2022)}]{ngampruetikorn:22}%
  \BibitemOpen
  \bibfield  {author} {\bibinfo {author} {\bibfnamefont {V.}~\bibnamefont
  {Ngampruetikorn}}\ and\ \bibinfo {author} {\bibfnamefont {D.~J.}\
  \bibnamefont {Schwab}},\ }\bibfield  {title} {\bibinfo {title} {Information
  bottleneck theory of high-dimensional regression: relevancy, efficiency and
  optimality},\ }in\ \href
  {https://proceedings.neurips.cc/paper_files/paper/2022/hash/3fbcfbc2b4009ae8dfa17a562532d123-Abstract-Conference.html}
  {\emph {\bibinfo {booktitle} {Advances in Neural Information Processing
  Systems}}},\ Vol.~\bibinfo {volume} {35},\ \bibinfo {editor} {edited by\
  \bibinfo {editor} {\bibfnamefont {S.}~\bibnamefont {Koyejo}}, \bibinfo
  {editor} {\bibfnamefont {S.}~\bibnamefont {Mohamed}}, \bibinfo {editor}
  {\bibfnamefont {A.}~\bibnamefont {Agarwal}}, \bibinfo {editor} {\bibfnamefont
  {D.}~\bibnamefont {Belgrave}}, \bibinfo {editor} {\bibfnamefont
  {K.}~\bibnamefont {Cho}},\ and\ \bibinfo {editor} {\bibfnamefont
  {A.}~\bibnamefont {Oh}}}\ (\bibinfo  {publisher} {Curran Associates, Inc.},\
  \bibinfo {year} {2022})\ pp.\ \bibinfo {pages} {9784--9796}\BibitemShut
  {NoStop}%
\bibitem [{\citenamefont {Chalk}\ \emph {et~al.}(2016)\citenamefont {Chalk},
  \citenamefont {Marre},\ and\ \citenamefont {Tkacik}}]{chalk:16}%
  \BibitemOpen
  \bibfield  {author} {\bibinfo {author} {\bibfnamefont {M.}~\bibnamefont
  {Chalk}}, \bibinfo {author} {\bibfnamefont {O.}~\bibnamefont {Marre}},\ and\
  \bibinfo {author} {\bibfnamefont {G.}~\bibnamefont {Tkacik}},\ }\bibfield
  {title} {\bibinfo {title} {Relevant sparse codes with variational information
  bottleneck},\ }in\ \href
  {https://proceedings.neurips.cc/paper/2016/hash/a89cf525e1d9f04d16ce31165e139a4b-Abstract.html}
  {\emph {\bibinfo {booktitle} {Advances in Neural Information Processing
  Systems}}},\ Vol.~\bibinfo {volume} {29},\ \bibinfo {editor} {edited by\
  \bibinfo {editor} {\bibfnamefont {D.~D.}\ \bibnamefont {Lee}}, \bibinfo
  {editor} {\bibfnamefont {M.}~\bibnamefont {Sugiyama}}, \bibinfo {editor}
  {\bibfnamefont {U.~V.}\ \bibnamefont {Luxburg}}, \bibinfo {editor}
  {\bibfnamefont {I.}~\bibnamefont {Guyon}},\ and\ \bibinfo {editor}
  {\bibfnamefont {R.}~\bibnamefont {Garnett}}}\ (\bibinfo  {publisher} {Curran
  Associates, Inc.},\ \bibinfo {year} {2016})\ pp.\ \bibinfo {pages}
  {1957--1965}\BibitemShut {NoStop}%
\bibitem [{\citenamefont {Poole}\ \emph {et~al.}(2019)\citenamefont {Poole},
  \citenamefont {Ozair}, \citenamefont {Van Den~Oord}, \citenamefont {Alemi},\
  and\ \citenamefont {Tucker}}]{poole:19}%
  \BibitemOpen
  \bibfield  {author} {\bibinfo {author} {\bibfnamefont {B.}~\bibnamefont
  {Poole}}, \bibinfo {author} {\bibfnamefont {S.}~\bibnamefont {Ozair}},
  \bibinfo {author} {\bibfnamefont {A.}~\bibnamefont {Van Den~Oord}}, \bibinfo
  {author} {\bibfnamefont {A.}~\bibnamefont {Alemi}},\ and\ \bibinfo {author}
  {\bibfnamefont {G.}~\bibnamefont {Tucker}},\ }\bibfield  {title} {\bibinfo
  {title} {On variational bounds of mutual information},\ }in\ \href
  {https://proceedings.mlr.press/v97/poole19a.html} {\emph {\bibinfo
  {booktitle} {Proceedings of the 36th International Conference on Machine
  Learning}}},\ \bibinfo {series} {Proceedings of Machine Learning Research},
  Vol.~\bibinfo {volume} {97},\ \bibinfo {editor} {edited by\ \bibinfo {editor}
  {\bibfnamefont {K.}~\bibnamefont {Chaudhuri}}\ and\ \bibinfo {editor}
  {\bibfnamefont {R.}~\bibnamefont {Salakhutdinov}}}\ (\bibinfo  {publisher}
  {PMLR},\ \bibinfo {year} {2019})\ pp.\ \bibinfo {pages}
  {5171--5180}\BibitemShut {NoStop}%
\bibitem [{\citenamefont {Tschannen}\ \emph {et~al.}(2020)\citenamefont
  {Tschannen}, \citenamefont {Djolonga}, \citenamefont {Rubenstein},
  \citenamefont {Gelly},\ and\ \citenamefont {Lucic}}]{tschannen:20}%
  \BibitemOpen
  \bibfield  {author} {\bibinfo {author} {\bibfnamefont {M.}~\bibnamefont
  {Tschannen}}, \bibinfo {author} {\bibfnamefont {J.}~\bibnamefont {Djolonga}},
  \bibinfo {author} {\bibfnamefont {P.~K.}\ \bibnamefont {Rubenstein}},
  \bibinfo {author} {\bibfnamefont {S.}~\bibnamefont {Gelly}},\ and\ \bibinfo
  {author} {\bibfnamefont {M.}~\bibnamefont {Lucic}},\ }\bibfield  {title}
  {\bibinfo {title} {On mutual information maximization for representation
  learning},\ }in\ \href {https://openreview.net/forum?id=rkxoh24FPH} {\emph
  {\bibinfo {booktitle} {International Conference on Learning
  Representations}}}\ (\bibinfo {year} {2020})\BibitemShut {NoStop}%
\bibitem [{\citenamefont {Wu}\ \emph {et~al.}(2019)\citenamefont {Wu},
  \citenamefont {Fischer}, \citenamefont {Chuang},\ and\ \citenamefont
  {Tegmark}}]{wu:19}%
  \BibitemOpen
  \bibfield  {author} {\bibinfo {author} {\bibfnamefont {T.}~\bibnamefont
  {Wu}}, \bibinfo {author} {\bibfnamefont {I.}~\bibnamefont {Fischer}},
  \bibinfo {author} {\bibfnamefont {I.~L.}\ \bibnamefont {Chuang}},\ and\
  \bibinfo {author} {\bibfnamefont {M.}~\bibnamefont {Tegmark}},\ }\bibfield
  {title} {\bibinfo {title} {Learnability for the information bottleneck},\
  }\href {https://doi.org/10.3390/e21100924} {\bibfield  {journal} {\bibinfo
  {journal} {Entropy}\ }\textbf {\bibinfo {volume} {21}},\ \bibinfo {pages}
  {924} (\bibinfo {year} {2019})}\BibitemShut {NoStop}%
\bibitem [{\citenamefont {Ngampruetikorn}\ and\ \citenamefont
  {Schwab}(2021)}]{ngampruetikorn:21}%
  \BibitemOpen
  \bibfield  {author} {\bibinfo {author} {\bibfnamefont {V.}~\bibnamefont
  {Ngampruetikorn}}\ and\ \bibinfo {author} {\bibfnamefont {D.~J.}\
  \bibnamefont {Schwab}},\ }\bibfield  {title} {\bibinfo {title} {Perturbation
  theory for the information bottleneck},\ }in\ \href
  {https://proceedings.neurips.cc/paper/2021/hash/af8d9c4e238c63fb074b44eb6aed80ae-Abstract.html}
  {\emph {\bibinfo {booktitle} {Advances in Neural Information Processing
  Systems}}},\ Vol.~\bibinfo {volume} {34},\ \bibinfo {editor} {edited by\
  \bibinfo {editor} {\bibfnamefont {M.}~\bibnamefont {Ranzato}}, \bibinfo
  {editor} {\bibfnamefont {A.}~\bibnamefont {Beygelzimer}}, \bibinfo {editor}
  {\bibfnamefont {Y.}~\bibnamefont {Dauphin}}, \bibinfo {editor} {\bibfnamefont
  {P.}~\bibnamefont {Liang}},\ and\ \bibinfo {editor} {\bibfnamefont {J.~W.}\
  \bibnamefont {Vaughan}}}\ (\bibinfo  {publisher} {Curran Associates, Inc.},\
  \bibinfo {year} {2021})\ pp.\ \bibinfo {pages} {21008--21018}\BibitemShut
  {NoStop}%
\bibitem [{\citenamefont {Chechik}\ \emph {et~al.}(2005)\citenamefont
  {Chechik}, \citenamefont {Globerson}, \citenamefont {Tishby},\ and\
  \citenamefont {Weiss}}]{chechik:05}%
  \BibitemOpen
  \bibfield  {author} {\bibinfo {author} {\bibfnamefont {G.}~\bibnamefont
  {Chechik}}, \bibinfo {author} {\bibfnamefont {A.}~\bibnamefont {Globerson}},
  \bibinfo {author} {\bibfnamefont {N.}~\bibnamefont {Tishby}},\ and\ \bibinfo
  {author} {\bibfnamefont {Y.}~\bibnamefont {Weiss}},\ }\bibfield  {title}
  {\bibinfo {title} {Information bottleneck for {G}aussian variables},\ }\href
  {https://www.jmlr.org/papers/v6/chechik05a.html} {\bibfield  {journal}
  {\bibinfo  {journal} {Journal of Machine Learning Research}\ }\textbf
  {\bibinfo {volume} {6}},\ \bibinfo {pages} {165} (\bibinfo {year}
  {2005})}\BibitemShut {NoStop}%
\bibitem [{\citenamefont {R{\'e}nyi}(1961)}]{renyi:61}%
  \BibitemOpen
  \bibfield  {author} {\bibinfo {author} {\bibfnamefont {A.}~\bibnamefont
  {R{\'e}nyi}},\ }\bibfield  {title} {\bibinfo {title} {On measures of entropy
  and information},\ }in\ \href@noop {} {\emph {\bibinfo {booktitle}
  {Proceedings of the Fourth Berkeley Symposium on Mathematical Statistics and
  Probability}}},\ Vol.~\bibinfo {volume} {1},\ \bibinfo {editor} {edited by\
  \bibinfo {editor} {\bibfnamefont {J.}~\bibnamefont {Neyman}}}\ (\bibinfo
  {year} {1961})\ pp.\ \bibinfo {pages} {547--561}\BibitemShut {NoStop}%
\bibitem [{\citenamefont {Horodecki}\ \emph {et~al.}(2009)\citenamefont
  {Horodecki}, \citenamefont {Horodecki}, \citenamefont {Horodecki},\ and\
  \citenamefont {Horodecki}}]{horodecki:09}%
  \BibitemOpen
  \bibfield  {author} {\bibinfo {author} {\bibfnamefont {R.}~\bibnamefont
  {Horodecki}}, \bibinfo {author} {\bibfnamefont {P.}~\bibnamefont
  {Horodecki}}, \bibinfo {author} {\bibfnamefont {M.}~\bibnamefont
  {Horodecki}},\ and\ \bibinfo {author} {\bibfnamefont {K.}~\bibnamefont
  {Horodecki}},\ }\bibfield  {title} {\bibinfo {title} {Quantum entanglement},\
  }\href {https://doi.org/10.1103/RevModPhys.81.865} {\bibfield  {journal}
  {\bibinfo  {journal} {Reviews of Modern Physics}\ }\textbf {\bibinfo {volume}
  {81}},\ \bibinfo {pages} {865} (\bibinfo {year} {2009})}\BibitemShut
  {NoStop}%
\bibitem [{\citenamefont {Eisert}\ \emph {et~al.}(2010)\citenamefont {Eisert},
  \citenamefont {Cramer},\ and\ \citenamefont {Plenio}}]{eisert:10}%
  \BibitemOpen
  \bibfield  {author} {\bibinfo {author} {\bibfnamefont {J.}~\bibnamefont
  {Eisert}}, \bibinfo {author} {\bibfnamefont {M.}~\bibnamefont {Cramer}},\
  and\ \bibinfo {author} {\bibfnamefont {M.~B.}\ \bibnamefont {Plenio}},\
  }\bibfield  {title} {\bibinfo {title} {Colloquium: Area laws for the
  entanglement entropy},\ }\href {https://doi.org/10.1103/RevModPhys.82.277}
  {\bibfield  {journal} {\bibinfo  {journal} {Reviews of Modern Physics}\
  }\textbf {\bibinfo {volume} {82}},\ \bibinfo {pages} {277} (\bibinfo {year}
  {2010})}\BibitemShut {NoStop}%
\bibitem [{\citenamefont {Hastings}\ \emph {et~al.}(2010)\citenamefont
  {Hastings}, \citenamefont {Gonz\'alez}, \citenamefont {Kallin},\ and\
  \citenamefont {Melko}}]{hastings:10}%
  \BibitemOpen
  \bibfield  {author} {\bibinfo {author} {\bibfnamefont {M.~B.}\ \bibnamefont
  {Hastings}}, \bibinfo {author} {\bibfnamefont {I.}~\bibnamefont
  {Gonz\'alez}}, \bibinfo {author} {\bibfnamefont {A.~B.}\ \bibnamefont
  {Kallin}},\ and\ \bibinfo {author} {\bibfnamefont {R.~G.}\ \bibnamefont
  {Melko}},\ }\bibfield  {title} {\bibinfo {title} {Measuring renyi
  entanglement entropy in quantum monte carlo simulations},\ }\href
  {https://doi.org/10.1103/PhysRevLett.104.157201} {\bibfield  {journal}
  {\bibinfo  {journal} {Physical Review Letters}\ }\textbf {\bibinfo {volume}
  {104}},\ \bibinfo {pages} {157201} (\bibinfo {year} {2010})}\BibitemShut
  {NoStop}%
\bibitem [{\citenamefont {Singh}\ \emph {et~al.}(2011)\citenamefont {Singh},
  \citenamefont {Hastings}, \citenamefont {Kallin},\ and\ \citenamefont
  {Melko}}]{singh:11}%
  \BibitemOpen
  \bibfield  {author} {\bibinfo {author} {\bibfnamefont {R.~R.~P.}\
  \bibnamefont {Singh}}, \bibinfo {author} {\bibfnamefont {M.~B.}\ \bibnamefont
  {Hastings}}, \bibinfo {author} {\bibfnamefont {A.~B.}\ \bibnamefont
  {Kallin}},\ and\ \bibinfo {author} {\bibfnamefont {R.~G.}\ \bibnamefont
  {Melko}},\ }\bibfield  {title} {\bibinfo {title} {Finite-temperature critical
  behavior of mutual information},\ }\href
  {https://doi.org/10.1103/PhysRevLett.106.135701} {\bibfield  {journal}
  {\bibinfo  {journal} {Physical Review Letters}\ }\textbf {\bibinfo {volume}
  {106}},\ \bibinfo {pages} {135701} (\bibinfo {year} {2011})}\BibitemShut
  {NoStop}%
\bibitem [{\citenamefont {Herdman}\ \emph {et~al.}(2017)\citenamefont
  {Herdman}, \citenamefont {Roy}, \citenamefont {Melko},\ and\ \citenamefont
  {Maestro}}]{herdman:17}%
  \BibitemOpen
  \bibfield  {author} {\bibinfo {author} {\bibfnamefont {C.~M.}\ \bibnamefont
  {Herdman}}, \bibinfo {author} {\bibfnamefont {P.~N.}\ \bibnamefont {Roy}},
  \bibinfo {author} {\bibfnamefont {R.~G.}\ \bibnamefont {Melko}},\ and\
  \bibinfo {author} {\bibfnamefont {A.~D.}\ \bibnamefont {Maestro}},\
  }\bibfield  {title} {\bibinfo {title} {Entanglement area law in superfluid
  \textsuperscript{4}{He}},\ }\href {https://doi.org/10.1038/nphys4075}
  {\bibfield  {journal} {\bibinfo  {journal} {Nature Physics}\ }\textbf
  {\bibinfo {volume} {13}},\ \bibinfo {pages} {556} (\bibinfo {year}
  {2017})}\BibitemShut {NoStop}%
\bibitem [{\citenamefont {Islam}\ \emph {et~al.}(2015)\citenamefont {Islam},
  \citenamefont {Ma}, \citenamefont {Preiss}, \citenamefont {Eric~Tai},
  \citenamefont {Lukin}, \citenamefont {Rispoli},\ and\ \citenamefont
  {Greiner}}]{islam:15}%
  \BibitemOpen
  \bibfield  {author} {\bibinfo {author} {\bibfnamefont {R.}~\bibnamefont
  {Islam}}, \bibinfo {author} {\bibfnamefont {R.}~\bibnamefont {Ma}}, \bibinfo
  {author} {\bibfnamefont {P.~M.}\ \bibnamefont {Preiss}}, \bibinfo {author}
  {\bibfnamefont {M.}~\bibnamefont {Eric~Tai}}, \bibinfo {author}
  {\bibfnamefont {A.}~\bibnamefont {Lukin}}, \bibinfo {author} {\bibfnamefont
  {M.}~\bibnamefont {Rispoli}},\ and\ \bibinfo {author} {\bibfnamefont
  {M.}~\bibnamefont {Greiner}},\ }\bibfield  {title} {\bibinfo {title}
  {Measuring entanglement entropy in a quantum many-body system},\ }\href
  {https://doi.org/10.1038/nature15750} {\bibfield  {journal} {\bibinfo
  {journal} {Nature}\ }\textbf {\bibinfo {volume} {528}},\ \bibinfo {pages}
  {77} (\bibinfo {year} {2015})}\BibitemShut {NoStop}%
\bibitem [{\citenamefont {Bergschneider}\ \emph {et~al.}(2019)\citenamefont
  {Bergschneider}, \citenamefont {Klinkhamer}, \citenamefont {Becher},
  \citenamefont {Klemt}, \citenamefont {Palm}, \citenamefont {Z{\"u}rn},
  \citenamefont {Jochim},\ and\ \citenamefont {Preiss}}]{bergschneider:19}%
  \BibitemOpen
  \bibfield  {author} {\bibinfo {author} {\bibfnamefont {A.}~\bibnamefont
  {Bergschneider}}, \bibinfo {author} {\bibfnamefont {V.~M.}\ \bibnamefont
  {Klinkhamer}}, \bibinfo {author} {\bibfnamefont {J.~H.}\ \bibnamefont
  {Becher}}, \bibinfo {author} {\bibfnamefont {R.}~\bibnamefont {Klemt}},
  \bibinfo {author} {\bibfnamefont {L.}~\bibnamefont {Palm}}, \bibinfo {author}
  {\bibfnamefont {G.}~\bibnamefont {Z{\"u}rn}}, \bibinfo {author}
  {\bibfnamefont {S.}~\bibnamefont {Jochim}},\ and\ \bibinfo {author}
  {\bibfnamefont {P.~M.}\ \bibnamefont {Preiss}},\ }\bibfield  {title}
  {\bibinfo {title} {Experimental characterization of two-particle entanglement
  through position and momentum correlations},\ }\href
  {https://doi.org/10.1038/s41567-019-0508-6} {\bibfield  {journal} {\bibinfo
  {journal} {Nature Physics}\ }\textbf {\bibinfo {volume} {15}},\ \bibinfo
  {pages} {640} (\bibinfo {year} {2019})}\BibitemShut {NoStop}%
\bibitem [{\citenamefont {Brydges}\ \emph {et~al.}(2019)\citenamefont
  {Brydges}, \citenamefont {Elben}, \citenamefont {Jurcevic}, \citenamefont
  {Vermersch}, \citenamefont {Maier}, \citenamefont {Lanyon}, \citenamefont
  {Zoller}, \citenamefont {Blatt},\ and\ \citenamefont {Roos}}]{brydges:19}%
  \BibitemOpen
  \bibfield  {author} {\bibinfo {author} {\bibfnamefont {T.}~\bibnamefont
  {Brydges}}, \bibinfo {author} {\bibfnamefont {A.}~\bibnamefont {Elben}},
  \bibinfo {author} {\bibfnamefont {P.}~\bibnamefont {Jurcevic}}, \bibinfo
  {author} {\bibfnamefont {B.}~\bibnamefont {Vermersch}}, \bibinfo {author}
  {\bibfnamefont {C.}~\bibnamefont {Maier}}, \bibinfo {author} {\bibfnamefont
  {B.~P.}\ \bibnamefont {Lanyon}}, \bibinfo {author} {\bibfnamefont
  {P.}~\bibnamefont {Zoller}}, \bibinfo {author} {\bibfnamefont
  {R.}~\bibnamefont {Blatt}},\ and\ \bibinfo {author} {\bibfnamefont {C.~F.}\
  \bibnamefont {Roos}},\ }\bibfield  {title} {\bibinfo {title} {Probing
  {R}{\'e}nyi entanglement entropy via randomized measurements},\ }\href
  {https://doi.org/10.1126/science.aau4963} {\bibfield  {journal} {\bibinfo
  {journal} {Science}\ }\textbf {\bibinfo {volume} {364}},\ \bibinfo {pages}
  {260} (\bibinfo {year} {2019})}\BibitemShut {NoStop}%
\bibitem [{\citenamefont {Jeffreys}(1946)}]{jeffreys:46}%
  \BibitemOpen
  \bibfield  {author} {\bibinfo {author} {\bibfnamefont {H.}~\bibnamefont
  {Jeffreys}},\ }\bibfield  {title} {\bibinfo {title} {An invariant form for
  the prior probability in estimation problems},\ }\href
  {https://doi.org/10.1098/rspa.1946.0056} {\bibfield  {journal} {\bibinfo
  {journal} {Proceedings of the Royal Society of London. Series A. Mathematical
  and Physical Sciences}\ }\textbf {\bibinfo {volume} {186}},\ \bibinfo {pages}
  {453} (\bibinfo {year} {1946})}\BibitemShut {NoStop}%
\bibitem [{\citenamefont {Aminian}\ \emph {et~al.}(2021)\citenamefont
  {Aminian}, \citenamefont {Bu}, \citenamefont {Toni}, \citenamefont
  {Rodrigues},\ and\ \citenamefont {Wornell}}]{aminian:21}%
  \BibitemOpen
  \bibfield  {author} {\bibinfo {author} {\bibfnamefont {G.}~\bibnamefont
  {Aminian}}, \bibinfo {author} {\bibfnamefont {Y.}~\bibnamefont {Bu}},
  \bibinfo {author} {\bibfnamefont {L.}~\bibnamefont {Toni}}, \bibinfo {author}
  {\bibfnamefont {M.}~\bibnamefont {Rodrigues}},\ and\ \bibinfo {author}
  {\bibfnamefont {G.}~\bibnamefont {Wornell}},\ }\bibfield  {title} {\bibinfo
  {title} {An exact characterization of the generalization error for the gibbs
  algorithm},\ }in\ \href
  {https://proceedings.neurips.cc/paper/2021/hash/445e24b5f22cacb9d51a837c10e91a3f-Abstract.html}
  {\emph {\bibinfo {booktitle} {Advances in Neural Information Processing
  Systems}}},\ Vol.~\bibinfo {volume} {34},\ \bibinfo {editor} {edited by\
  \bibinfo {editor} {\bibfnamefont {M.}~\bibnamefont {Ranzato}}, \bibinfo
  {editor} {\bibfnamefont {A.}~\bibnamefont {Beygelzimer}}, \bibinfo {editor}
  {\bibfnamefont {Y.}~\bibnamefont {Dauphin}}, \bibinfo {editor} {\bibfnamefont
  {P.}~\bibnamefont {Liang}},\ and\ \bibinfo {editor} {\bibfnamefont {J.~W.}\
  \bibnamefont {Vaughan}}}\ (\bibinfo  {publisher} {Curran Associates, Inc.},\
  \bibinfo {year} {2021})\ pp.\ \bibinfo {pages} {8106--8118}\BibitemShut
  {NoStop}%
\bibitem [{\citenamefont {van Erven}\ and\ \citenamefont
  {HarremoÃ«s}(2014)}]{vanerven:14}%
  \BibitemOpen
  \bibfield  {author} {\bibinfo {author} {\bibfnamefont {T.}~\bibnamefont {van
  Erven}}\ and\ \bibinfo {author} {\bibfnamefont {P.}~\bibnamefont
  {HarremoÃ«s}},\ }\bibfield  {title} {\bibinfo {title} {RÃ©nyi divergence and
  kullback-leibler divergence},\ }\href
  {https://doi.org/10.1109/TIT.2014.2320500} {\bibfield  {journal} {\bibinfo
  {journal} {IEEE Transactions on Information Theory}\ }\textbf {\bibinfo
  {volume} {60}},\ \bibinfo {pages} {3797} (\bibinfo {year}
  {2014})}\BibitemShut {NoStop}%
\bibitem [{\citenamefont {Liese}\ and\ \citenamefont {Vajda}(2006)}]{liese:06}%
  \BibitemOpen
  \bibfield  {author} {\bibinfo {author} {\bibfnamefont {F.}~\bibnamefont
  {Liese}}\ and\ \bibinfo {author} {\bibfnamefont {I.}~\bibnamefont {Vajda}},\
  }\bibfield  {title} {\bibinfo {title} {On divergences and informations in
  statistics and information theory},\ }\href
  {https://doi.org/10.1109/TIT.2006.881731} {\bibfield  {journal} {\bibinfo
  {journal} {IEEE Transactions on Information Theory}\ }\textbf {\bibinfo
  {volume} {52}},\ \bibinfo {pages} {4394} (\bibinfo {year}
  {2006})}\BibitemShut {NoStop}%
\end{thebibliography}
%

\end{document}